\UseRawInputEncoding
\documentclass[reprint,
  amsmath,amssymb,
  aps,
  prb, superscriptaddress            % change to prb, prc, prd, pre, prl as needed
]{revtex4-2}

\usepackage{graphicx}  % figures
\usepackage{dcolumn}   % aligned columns
\usepackage{bm}        % bold math
\usepackage{hyperref}  % hyperlinks
\usepackage{tikz}
\usepackage{braket}
\usepackage{cleveref}

\newcommand{\Red}[1]{\textcolor{black}{#1}}

\begin{document}

\title{Entangled photons from quantum-dot--cavity systems under non-Markovian decoherence by pulsed excitation}

\author{Katy Snow}
\email{ksnow01@qub.ac.uk}
\affiliation{Centre for Quantum Materials and Technologies, School of Mathematics and Physics, Queen’s University Belfast, BT7 1NN, United Kingdom}
\author{Mauro Paternostro}
\affiliation{Universit\`{a} degli Studi di Palermo, Dipartimento di Fisica e Chimica - Emilio Segr\`{e}, via Archirafi 36, I-90123 Palermo, Italy}
\affiliation{Centre for Quantum Materials and Technologies, School of Mathematics and Physics, Queen’s University Belfast, BT7 1NN, United Kingdom}

\date{\today}

\begin{abstract}
Cascaded emission from the biexciton state of a quantum dot results in polarization entangled photon pairs. \Red{Cavity-enhancement of the direct two-photon emission channel bypasses the dominant source of decoherence in this system -- that due to fine-structure splitting of the exciton levels. Here, we investigate the remaining non-Markovian phonon-induced decoherence using the numerically exact uniTEMPO algorithm.} We compute the degree of entanglement of photon pairs generated by pulsed two-photon resonant excitation and find surprisingly good agreement between the numerically exact results and those calculated using the approximate polaron master equation method, permitting an efficient exploration of trends across system parameters.
\end{abstract}

\maketitle

\section{Introduction}
Non-classical states of light are essential for quantum communication, with entangled photon pairs being a paradigmatic example and the basis of many protocols. These can be generated deterministically using semiconductor quantum dots (QDs). Recent experiments have demonstrated quantum teleportation and all-photonic entanglement swapping using photon pairs from remote quantum dots~\cite{Laneve2025, Beccaceci2025arxiv}~\Red{\cite{Strobel2025}} - both protocols being primitives for quantum repeaters, the essential component in long-distance transfer of quantum information~\cite{Azure2023}.

In the standard experimental configuration, the two entangled photons are emitted sequentially from the biexciton state of the QD via the exciton states, with fine-structure splitting of the exciton states giving rise to which-path information and being the limiting factor in the degree of entanglement of emitted photons~\cite{Pfanner2008}. Cavity-enhancement of the direct two-photon transition was proposed as a method of avoiding the intermediate exciton states and thus increasing the entanglement~\cite{delValle2011}. This has been shown to hold even in the presence of non-Markovian phonon-induced decoherence~\cite{Heinz2017}, and has been demonstrated experimentally~\cite{Liu2025}. \Red{This unconventional source of quantum correlations emits frequency-degenerate photon pairs with hyperentanglement -- they are strongly correlated in both the polarization and time-bin degrees of freedom~\cite{delValle2013, Liu2025, Prilmuller2018}. This temporal correlation limits the applicability to standard protocols, and can be eliminated by enhancing the emission rate of the first photon~\cite{Simon2005, Fischer2016, Gustin2018, Scholl2020, Bauch2023, Heinisch2026}.}

The final limiting factor on the degree of entanglement is imperfect initialization of the system in the biexciton state. In the absence of cavity-mediated transition, finite laser excitation pulse widths have a detrimental effect due to Stark-shifting of the intermediate exciton states~\cite{Seidelmann2022_pulsed, Basset2023}, and this effect becomes more pronounced in the presence of a phonon environment~\cite{Seidelmann2023_pulsed_phonon}. However, as Stark-shifting of the exciton states will not have the same relevance, it is unclear what the combined effect of phonons and finite pulse widths in the cavity-mediated case will be. \Red{It was found that a two-colour excitation scheme can give very high entanglement compared to the resonant scheme, even at elevated temperatures~\cite{Bracht2023}.} In a driven QD-cavity system the multitude of processes occurring in the dressed state basis often make an intuitive understanding and associated approximate analytic results difficult, and demand a full calculation of the dynamics~\cite{Seidelmann2023_const_drive, delValle2010}. \Red{Inclusion of both the phonon environment and pulsed excitation in a numerical model of the dynamics can be achieved using an approximate polaron master equation~\cite{Gustin2018, Bauch2021, Bauch2023, Dewan2026, Heinisch2026}, with more recent works making use of tensor network methods in order to obtain exact results~\cite{Bracht2023, Liu2025}.}

\Red{Earlier numerically exact studies~\cite{Cygorek2018_biex, Seidelmann2023_const_drive, Seidelmann2023_pulsed_phonon, Glassl2012} of the non-Markovian phonon-induced decoherence of exciton-biexciton systems were performed using the quasi-adiabatic path-integral (QUAPI),} a real-time path-integral method~\cite{Makri1995a, Makri1995b}. Recent tensor network methods have developed this method further, with time-evolving matrix product operators (TEMPO) incorporating tensor network compression techniques~\cite{Strathearn2018}. TEMPO was subsequently reconciled with the process tensor formalism~\cite{Pollock2018}, resulting in a more powerful description of the non-Markovian dynamics, besides a greater computational efficiency~\cite{Jorgensen2019}. A reformulation of the tensor network contraction sequence using infinite time-evolving block decimation, known as uniTEMPO, allows faster contraction and results in a time-translationally-symmetric process tensor~\cite{Link2024}.

Alternative numerically exact methods include the time evolving density matrix using orthogonal polynomials algorithm (TEDOPA)~\cite{Chin2010, Prior2010}, in which the full environmental dynamics are modeled, meaning that the environment doesn't limit the system size, but modeling long-time dynamics is computationally costly. Another method is the Trotter decomposition with linked cluster expansion technique~\cite{Morreau2019}, which may be scalable to larger system sizes but currently only describes Hamiltonian evolution without decay~\cite{Hall2025}.

\Red{Here, we make use of a TEMPO algorithm that exploits degeneracies in the spectrum of the phonon coupling operator to make the simulation of systems with large Hilbert spaces feasible~\cite{Cygorek2017, Strathearn2018, Fux2024, Cygorek2024-ACE}. We thus use this method to compute the dynamics of the QD-cavity-phonon system under pulsed resonant two-photon excitation, with cavity enhancement of the direct two-photon transition,} and calculate the degree of entanglement of emitted photon pairs. 

We also implement a polaron master equation, beyond the previously used weak-coupling approximation~\cite{Roy2011, Dewan2026}, and find good agreement of the calculated concurrence, a measure of the degree of entanglement of photon pairs, with the numerically exact result. The polaron master equation is a much more computationally efficient method and is subsequently used to calculate the dependence of the concurrence on driving pulse width and temperature. We find that finite pulse widths do still have an adverse effect on concurrence, even in spite of the inherent insensitivity of spontaneous two-photon emission to Stark shifting of the exciton states. Moreover, we find that this negative effect is compounded by phonon-induced decoherence.

\section{System}

\begin{figure}
  \includegraphics[width=\columnwidth]{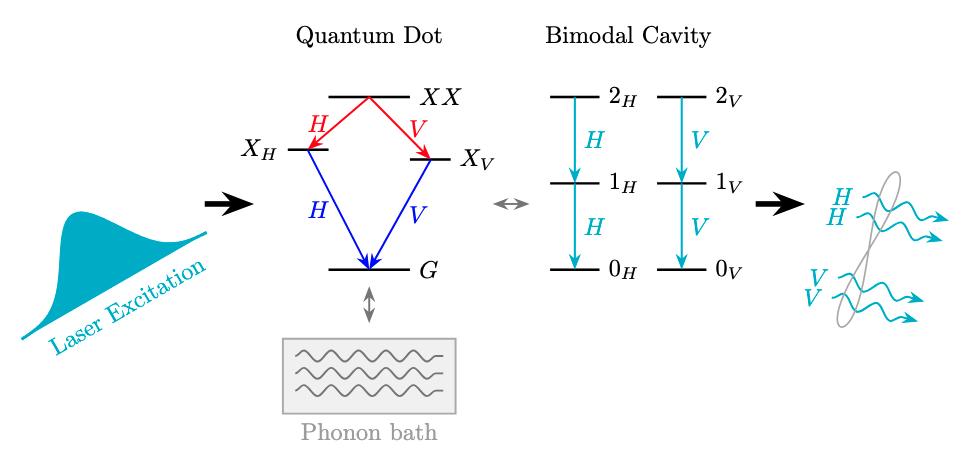}
  \caption{Schematic of the pulsed QD-cavity-phonon system under consideration. The quantum dot is excited from the ground $G$ to biexciton $XX$ state by two-photon excitation with a Gaussian laser pulse at frequency equal to half the biexciton frequency. Cascaded emission from $XX$ to $G$ via the exciton states $X_{H/V}$ results in polarization-entangled photon pairs. A bimodal cavity tuned to half the biexciton frequency enhances the direct two-photon emission process, meanwhile lattice vibrations in the substrate result in a non-Markovian decoherence.}
  \label{fig:1}
\end{figure}

The quantum dot is modelled as a 4-level system, as shown in Fig.~\ref{fig:1}, with ground state $|G\rangle$, exciton states $|X_H\rangle$ and $|X_V\rangle$ separated by a fine structure splitting $\delta_{\mathrm{fs}}$, and biexciton state $|XX\rangle$ with binding energy $E_B$. To enhance the direct two-photon emission channel, the QD is coupled to a cavity tuned to half the biexciton energy. We consider a bimodal cavity supporting both $H$- and $V$-polarised modes of equal frequency. Selection rules of the quantum dot demand that transition from the biexciton state to ground occurs either via the $|X_H\rangle$ state with emission of two $H$-polarised photons or via the $|X_V\rangle$ state with emission of two $V$-polarised photons, resulting, ideally, in a maximally entangled Bell state. The biexciton state is excited through two-photon excitation by a laser pulse at half the biexciton energy, resulting in the rotating-frame QD-cavity Hamiltonian $H_{\mathrm{S}} = H_{\mathrm{QD}} + H_{\mathrm{int}} + H_{\mathrm{drive}},$
where~\cite{Heinz2017, Seidelmann2023_const_drive}
\begin{equation}
\begin{aligned}
&H_{\mathrm{QD}}
=\left( \frac{E_B}{2} + \frac{\delta_{\mathrm{fs}}}{2} \right)\, |X_H\rangle\langle X_H|
%\\ &\qquad \qquad 
+ \left( \frac{E_B}{2} - \frac{\delta_{\mathrm{fs}}}{2} \right) \, |X_V\rangle\langle X_V|,\\
&H_{\mathrm{int}} = \: g \left(
\sigma_H a_H^\dagger
+ \sigma_V a_V^\dagger \right) \, + \, \mathrm{h.c.},\\
&H_{\mathrm{drive}}
= \: \frac{\Omega}{\sqrt{2}} \left(\sigma_H + \sigma_V \right) \, + \, \mathrm{h.c.},
\end{aligned}
\end{equation}
and $\sigma_i = |G\rangle\langle X_i| + |X_i\rangle\langle XX|, \: i \in \{H, V\}.$ Here, $g$ is the QD-cavity coupling strength, $a_{H/V}$ are the cavity annihilation operators, $\Omega$ is the driving strength, $H_{\mathrm{drive}}$ corresponds to diagonally polarised laser driving, and we consider Gaussian driving pulses of the form $\Omega(t) = \Omega_0 \exp\!\left(-(t-t_0)^2/2\tau^2\right)$. \Red{The polarisation of the driving field is not expected to have a strong impact on the resulting two-photon state~\cite{Seidelmann2023_pulsed_phonon}.} The cavity excitations subsequently decay at rate $\gamma$, which is included in the free system propagator $\mathcal{U}(t, t+\Delta t) = \exp{(\mathcal{L}(t) \Delta t)}$ as a Lindblad term in the Liouvillian~\cite{Heinz2017}
\begin{equation}
\mathcal{L}[\rho] = -i[H_S,\rho]
+ \gamma\sum_{k = H,V} 
\left(
a_k \rho a_k^\dagger
- \frac{1}{2}\{a_k^\dagger a_k,\rho\}
\right).
\end{equation}

As a measure of the entanglement of emitted photon pairs we use the time-dependent concurrence $C(t)$. This is proportional to the off-diagonal elements of the two-photon density matrix according to the expression
\begin{equation}
\begin{aligned}
C(t) & = 2 \, |\rho_{HV}(t)| =  \frac{2|G_{HV}(t)|}{G_{HH}(t) + G_{VV}(t)},
\end{aligned}
\end{equation}
where $G_{ij}(t) = \mathrm{Tr}\left[\rho(t) \, a_i^\dagger a_i^\dagger a_j a_j\right]$ and $\rho(t)$ is the density matrix of the full QD-cavity system. Experimentally, the time-dependent concurrence corresponds to the degree of entanglement of photons detected simultaneously at a given time $t$. In order to maximise the efficiency of an emitter of entangled photons, it is necessary to minimise the filtering in time of detection events. The two-time integrated concurrence is a measure of the entanglement when photon pairs are filtered neither in the time interval separating them nor in the time of detection of the first photon~\cite{Cygorek2018_biex}. This measure is, however, costly to compute as it involves calculating two-time correlation functions, and we therefore consider instead the single-time integrated concurrence, which represents an upper bound to the achievable entanglement in the absence of filtering. This is found by including all simultaneous detection events, $\bar{C} = 2 \, |\langle G_{HV} \rangle|/\left(\langle G_{HH} \rangle + \langle G_{VV} \rangle \right),$ where $\langle \cdot \rangle$ indicates time-averaging~\cite{Cygorek2018_biex}.

Coupling to longitudinal acoustic phonons is described by the spin-boson interaction Hamiltonian \cite{Glassl2012}
\begin{equation}
\label{eq:H_int}
    H_{\mathrm{QD-ph}} = \sum_k \omega_k b_k^\dagger b_k 
      + \sum_k \left( g_k b_k^\dagger + g_k^* b_k \right) \hat{O},
\end{equation}
where $\hat{O} = \sum_u \lambda_u \, |u\rangle \langle u|$ is the system coupling operator, with $\lambda_u = \{0, 1, 1, 2\}$ for $u = \{G, X_H, X_V, XX\}$, and the effect of the phonon environment is fully characterised by its spectral density $J(\omega) = \sum_k |g_k|^2 \delta(\omega - \omega_k)$. In the following we consider typical values for a GaAs quantum dot, with details of the spectral density given in Appendix~\ref{app:sd}.

\section{Tensor Network Methods}

\begin{figure}
  \includegraphics[width=\columnwidth]{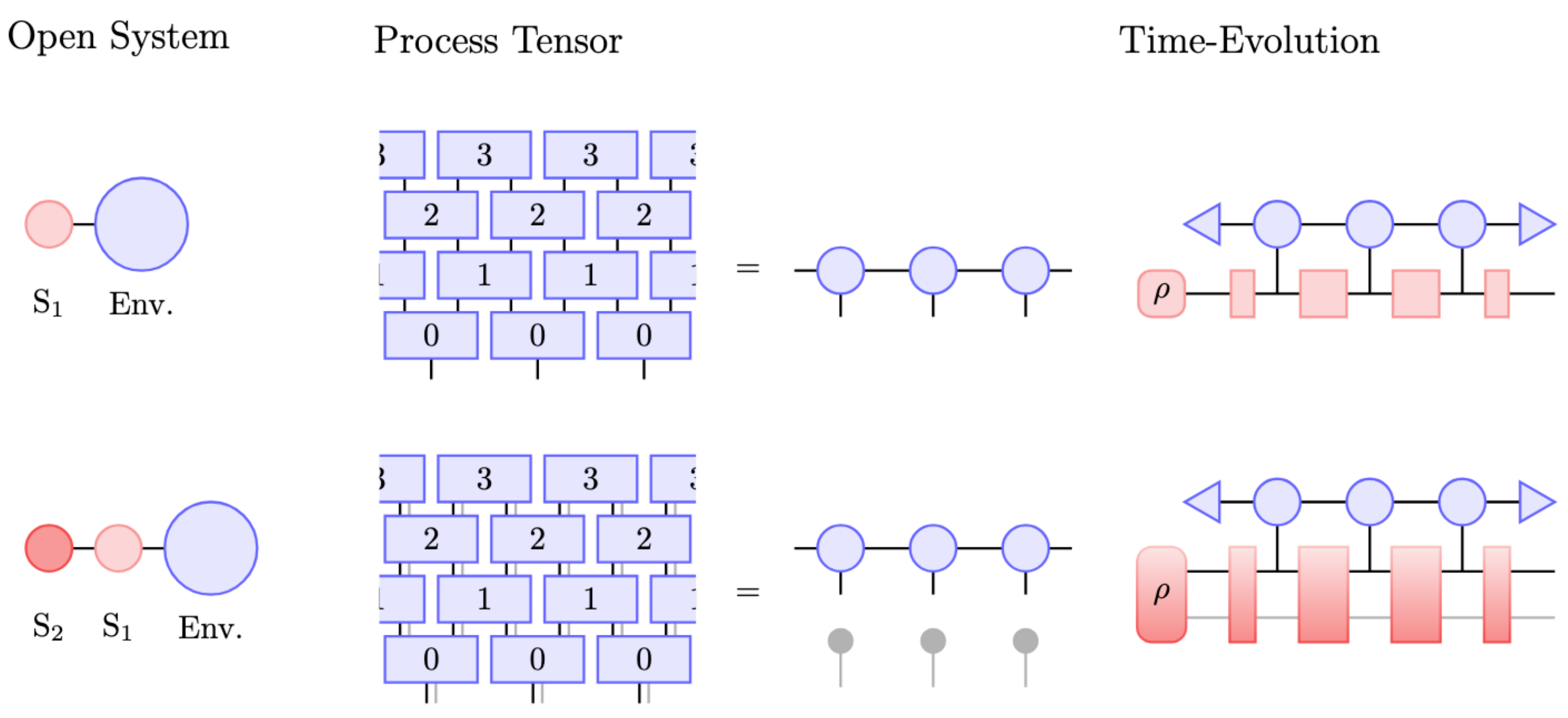}
  \caption{Illustration of the process tensor method for time-evolving the state of an open system under non-Markovian decoherence. In the standard scenario, shown in the first row, the system $S_1$ is coupled to an environment, and the influence of the environment is captured by a process tensor, which is constructed from an infinite network of elementary tensors (blue rectangles), and subsequently compressed and contracted (blue circles). This process tensor, alongside the free system propagator (pink rectangles), is then used to time-evolve the density matrix $\rho$, with finite-time boundary conditions imposed (blue triangles). In the second row we consider a system-environment coupling operator with degenerate eigenvalues. This corresponds to a system that can be divided into two subsystems; a subsystem $S_1$ that interacts with the environment and a subsystem $S_2$ that does not. The process tensor can then be decomposed into a smaller process tensor acting on the first subsystem (black tensor legs) and a part resulting in trivial evolution of the other subsystem (gray tensor legs).}
  \label{fig:2}
\end{figure}

Given a system-environment coupling as in Eq.~\eqref{eq:H_int}, and with a structured spectral density, the resulting non-Markovian dynamics of the system can be tracked using the numerically exact TEMPO method~\cite{Strathearn2018}. We consider the process tensor formulation of TEMPO~\cite{Jorgensen2019}, where the impact of the environment on the system is captured by a tensor $\mathcal{F}$ such that the density matrix of the system evolves over discrete time steps $\Delta t$ as
\begin{equation}
[\rho_n]_{\mu_n}
=
\mathcal{F}^{\mu_n,\ldots,\mu_1}
\left(
\prod_{l=1}^{n}
[\,\mathcal{U}_l\,]_{\mu_l,\mu_{l-1}}
\right)
[\rho_0]_{\mu_0},
\label{eq:2}
\end{equation}
where $[\rho_n]_{\mu}$ is the density matrix at time $n\,\Delta t$, with $\mu = (\mu_l, \mu_r)$ the index in Liouville notation such that $\rho_\mu=\bra{\mu_l}\rho\ket{\mu_r}$, and $\mathcal{U}_n = \mathcal{U}(n\Delta t, (n+1)\Delta t)$ is the free system propagator. \Red{The process tensor is computed using a discretized Feynman-Vernon path-integral, as outlined in Appendix B. We use the uniTEMPO algorithm~\cite{Link2024}, which results in the uniform process tensor shown in Fig.~\ref{fig:2}. The degeneracies in the spectrum of the phonon coupling operator are exploited, allowing us to use a minimal process tensor corresponding to the set of non-degenerate eigenvalues to fully capture the effect of the environment~\cite{Cygorek2017, Strathearn2018, Fux2024, Cygorek2024-ACE}, as depicted in the lower row of Fig.~\ref{fig:2}.}

For large system sizes the computational cost of determining  the dynamics is no-longer limited by the initial calculation of the process tensor, but instead by the propagation in time of the system density matrix. Using the Arnoldi algorithm in combination with the sparse matrix Liouvillian, the cost of time-evolution scales linearly with the internal bond dimension $\chi$ of the process tensor.

\section{Polaron Master Equation}
In order to obtain a more computationally tractable description of the open-system dynamics, we perform a polaron transformation and derive the corresponding polaron master equation (PME). This transformation is designed to eliminate the system-environment coupling term. The residual interaction is then taken to be a perturbation, and a master equation is derived in this polaron frame. We consider a generic system with Hamiltonian
\begin{equation}
    H_S = H_S(\bm{\delta}, \bm{g}) = \sum_i\delta_i \ket{i}\bra{i} + \sum_{i>j}g_{ij}\left(\ket{i}\bra{j}+\ket{j}\bra{i}\right)
\end{equation}
with $\delta_i,\,g_{ij}\in\mathbb{R}$. 
As for the interaction term, we assume it to be as in Eq.~\eqref{eq:H_int}. We then notice that, when written in this form, our previous system Hamiltonian satisfies $g_{ij}=0$ when $\lambda_i-\lambda_j \neq 1$, which means that system transitions only occur between states with adjacent phonon coupling eigenvalues. This means that the polaron master equation will be directly analogous to the two-level system case. We get~\cite{McCutcheon2011, Nazir2016}
\begin{equation}
\begin{aligned}
    \frac{d\rho}{dt} = & -i[H_S', \rho] - \sum_{u=(x, y)}\frac{1}{2}[X_u, D_u\rho - \rho D_u^\dagger]\\
    &- \sum_{u=(x, y)} i\, [X_u, S_u\rho + \rho S_u^\dagger],
\end{aligned}
\end{equation}
where $H_S' = H_S(\bm{\delta}', \bm{g}')$ with $\delta_i' = \delta_i-\lambda_i^2\Delta_p, \; g_{ij}' = \langle B \rangle g_{ij}$ where $\Delta_p$ is the polaron shift and $\langle B \rangle$ is the phonon-induced renormalization factor (see Appendix \ref{app:pme}). The operators $X_u, D_u$ and $S_u$ are given by $X_u = \sum_{i>j}g_{ij} \sigma_{(u)ij}$, with the $\sigma_{(u)}$ being the Pauli operators, and we have introduced the operators
\begin{equation}
\begin{aligned}
    &D_u = \sum_{ij}2\,\mathrm{Re}[\kappa_u(\Delta_{ij})] \, \bra{v_i}X_u \ket{v_j} \ket{v_i}\bra{v_j},\\ 
    &S_u = \sum_{ij}\mathrm{Im}[\kappa_u(\Delta_{ij})] \, \bra{v_i}X_u \ket{v_j} \ket{v_i}\bra{v_j},
\end{aligned}
\end{equation}
where $\ket{v_i}$ are the vectors that diagonalise the Hamiltonian $H_S' = \sum_i \omega_i \ket{v_i}\bra{v_i}$, and $\Delta_{ij}=\omega_i-\omega_j.$ The response functions $\kappa_u(\omega)$ depend on the bath correlation function, and the expression is given in Appendix \ref{app:pme}.

We highlight the fact that we have used a full Hamiltonian diagonalization approach rather than making the assumption of an approximately diagonal Hamiltonian, as has been done previously~\cite{Roy2011, Dewan2026}.

\section{Numerical Results}
\subsection{Initialization in the Biexciton State}

\begin{figure}
  \includegraphics[width=\linewidth]{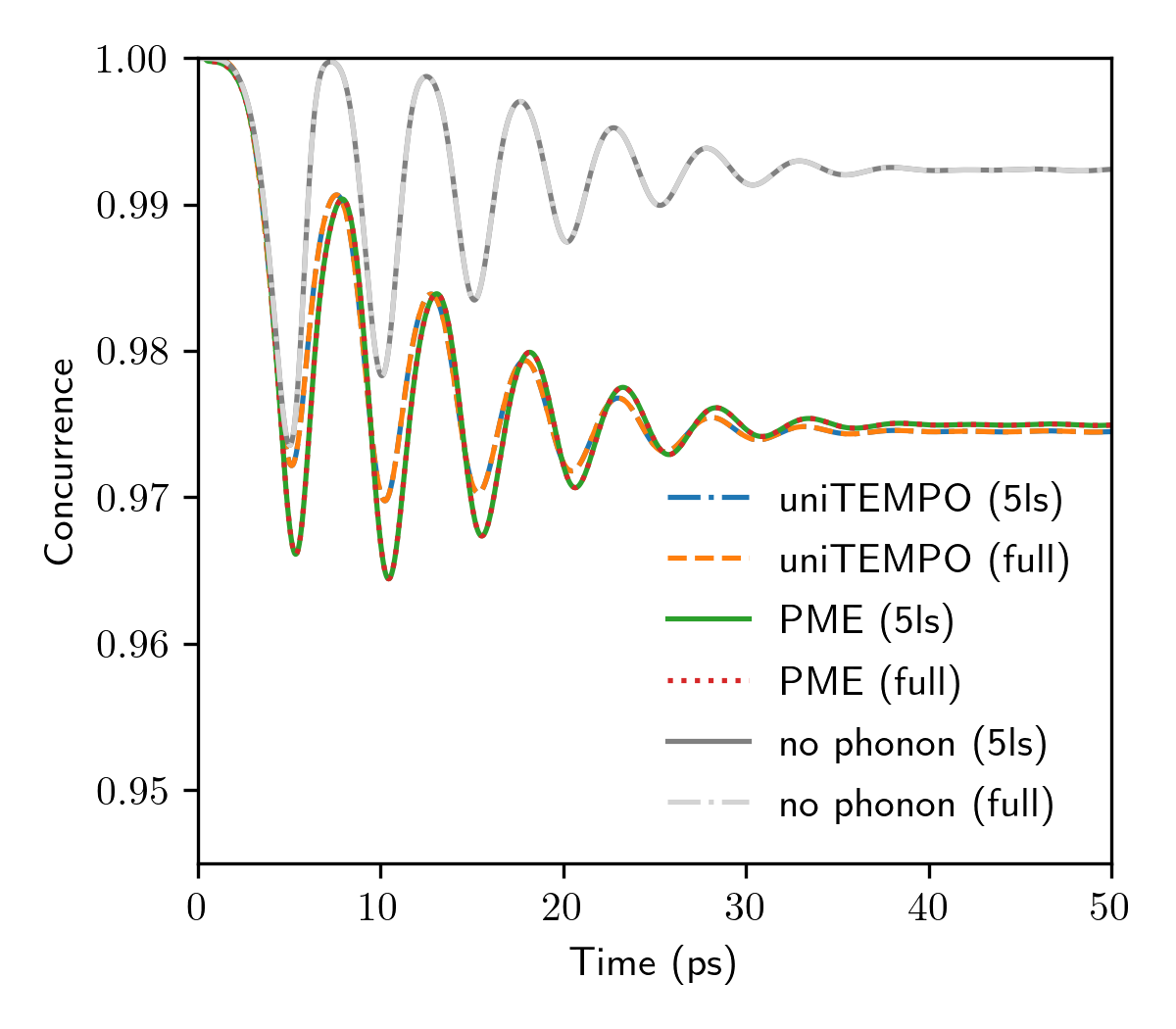}
  \caption{The degree of entanglement, measured by the concurrence, of photons emitted by the QD-cavity-phonon system following initialization in the biexciton ($XX$) state, as calculated using three different methods; the numerically exact uniTEMPO algorithm, the approximate polaron master equation (PME) and the case of no phonon environment. Calculations performed in the full Hilbert space are compared to those in the two-excitation subspace (denoted 5LS).}
  \label{fig:3}
\end{figure}

Firstly, in order to validate numerically our eigenvalue separation algorithm for the process tensor, we take the QD-cavity system to be initialised in $\ket{XX}$, with the cavity in the vacuum state, and allowed to decay with no external driving. In this special case the dynamics can be solved in the two-excitation subspace, which comprises 5 levels~\cite{Cygorek2018_biex}. Labelling the levels of the full system by $\ket{\nu, n_H, n_V}$, where $\nu$ is the state of the QD, and $n_{H/V}$ are the occupations of the two cavity modes, the two-excitation subspace consists of the levels $\ket{XX, 0, 0}, \ket{X_H, 1, 0}, \ket{X_V, 0, 1}, \ket{G,2,0}$ and $\ket{G,0,2}.$

The dynamics computed for the full Hilbert space can then be compared against those in the two-excitation subspace. For uniTEMPO calculations, we use a process tensor with external legs of dimension $3^2$ corresponding to the three eigenvalues $\lambda = \{0,1,2\}.$ We compare these numerically exact results to the dynamics computed using the polaron master equation, and to the dynamics in the absence of phonons, as shown in Fig.~\ref{fig:3}. Here and subsequently, we use the following set of parameters. The QD-cavity coupling $g$ and fine structure splitting $\delta_{\mathrm{fs}}$ are both equal to 0.1 meV, the biexciton binding energy $E_B$ is 1.5 meV, the cavity decay rate $\gamma$ is 0.25 ps$^{-1}$, corresponding to an intermediate coupling regime, and the phonon temperature is 10 K. \Red{High cavity coupling rates, as that chosen here, are of interest due to the role of the cavity in enhancing the direct two-photon emission channel, and prohibit a weak-coupling solution to the dynamics~\cite{Carmele2011}.}

As expected, the full space results agree with the 5-level system results in all cases, validating our process tensor time-evolution method. Furthermore, the time-dependent concurrence given by the PME agrees surprisingly well with the numerically exact result, especially at long times. \Red{The good agreement is attributed to the fact that the driving strengths considered here are weak compared to the characteristic energy scale of the phonon bath~\cite{McCutcheon2010}. It is important to note that the PME approach is unable to correctly capture multi-time correlations without a non-Markovian extension to the quantum regression theorem~\cite{McCutcheon2016}, and is known to systematically overestimate phonon-induced effects~\cite{Cosacchi2021}. The good agreement seen here is therefore specific to the single-time integrated concurrence used to measure the degree of entanglement, and will break down when looking at two-time integrated concurrences.}

\subsection{Pulsed Excitation}

\begin{figure}
  \includegraphics[width=\linewidth]{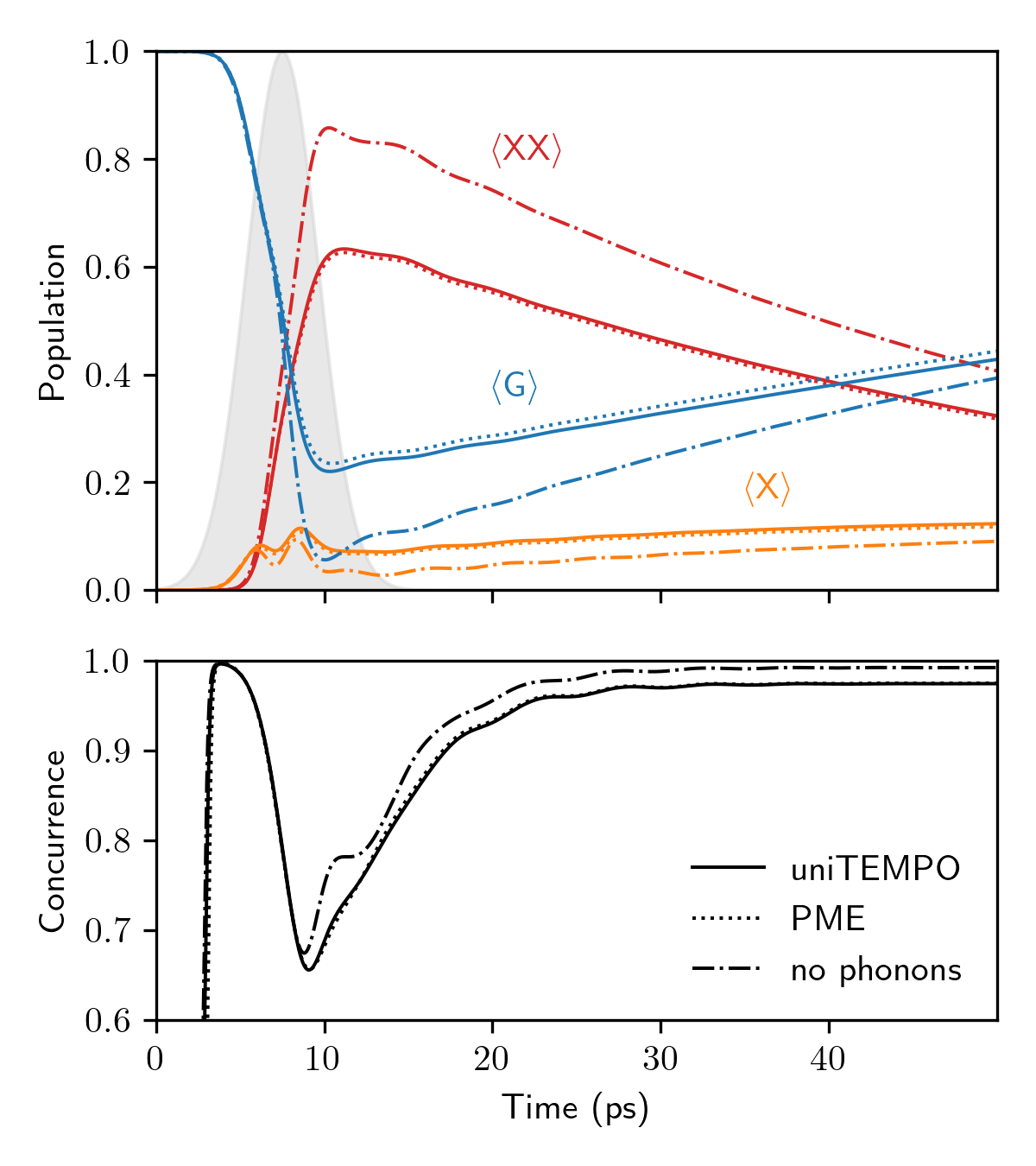}
  \caption{Dynamics of the QD-cavity-phonon system under pulsed two-photon excitation of the biexciton state from ground, calculated as before using three different methods; the numerically exact uniTEMPO, approximate PME, and the no-phonon case. Top: populations of the biexciton ($XX$), exciton ($X$), and ground ($G$) states, with the Gaussian laser pulse overlaid in gray. Bottom: concurrence is used as a measure of the entanglement of emitted photon pairs.}
  \label{fig:4}
\end{figure}

We next consider initialization in the ground state followed by pulsed excitation to the biexciton state and subsequent decay back down towards ground. The system populations and the time-dependent photon concurrence are shown in Fig.~\ref{fig:4} for a pulse of full-width-half-maximum $\Delta = 5.0$ ps and height $\Omega_0 = 0.545$ meV, as calculated using both uniTEMPO and the PME, alongside the no-phonon result.

The phonon environment has the effect of reducing the maximum population of the biexciton state, and therefore the efficiency of photon pair generation. Moreover, the pulse has a strongly negative effect on the photons' entanglement, with the concurrence recovering to it's previous level once the pulse duration is over. We also see that the concurrence is consistently lower in the presence of the phonon environment. Given that the PME agrees very well with the numerically exact result, we use this method to study these two effects in greater depth. 

\begin{figure}
  \includegraphics[width=\linewidth]{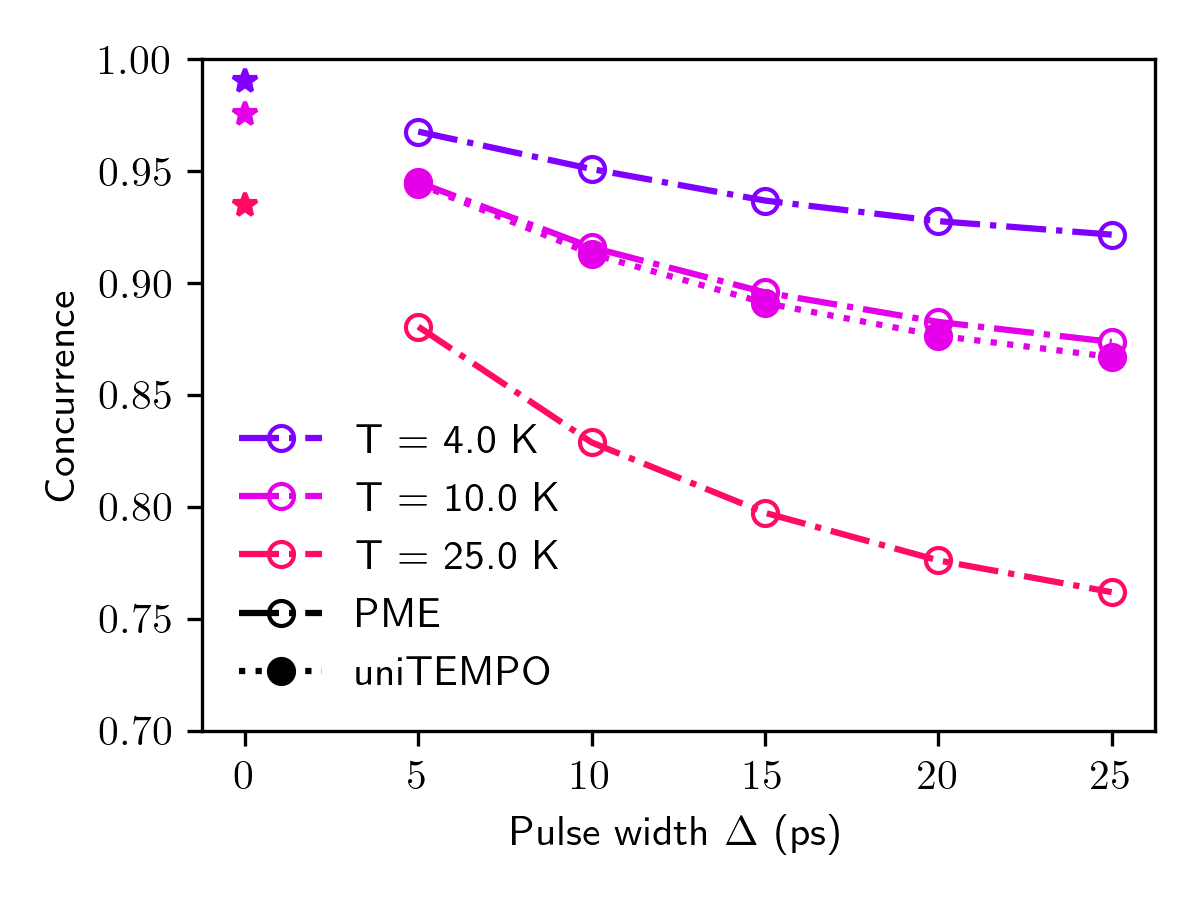}
  \caption{The concurrence of entangled photon pairs emitted by the QD-cavity-phonon system, following pulsed two-photon excitation of the biexciton state, is shown as a function of pulse width for three different phonon bath temperatures. The numerically exact uniTEMPO method is used to verify the approximate PME result for the 10~K temperature series. Also shown are the maximal concurrences for each temperature, which are computed for the system initialised in the biexciton state and represented by stars.}
  \label{fig:5}
\end{figure}

Fig.~\ref{fig:5} shows the dependence on pulse duration and phonon bath temperature of the single-time-integrated concurrence. When varying the pulse width, we also vary the pulse height such that the biexciton population is maximised after excitation. Additionally, we have computed the optimal, `zero pulse width', concurrence by initialising the system in $\ket{XX},$ and these results are represented by stars in Fig.~\ref{fig:5}. The numerically exact method is used to validate one of the temperature series’. \Red{We have checked that the PME dynamics for the remaining two temperatures also agree to a similar degree of accuracy.} We have absorbed polaron-induced shifts into the QD Hamiltonian such that the level splittings are temperature independent, to allow for easier comparison of results. \Red{Details of the required computational resources are listed in Appendix~\ref{app:comp}.}

It can be seen that, analogously to the no-cavity setup~\cite{Seidelmann2023_pulsed_phonon}, increasing the pulse width results in lower entanglement, whilst increasing the temperature compounds this effect. We thus conclude that, as in the cavity-less case, the negative effect of longer pulse durations outweighs the positive effect of weaker pulse strengths. The precise underlying mechanisms responsible for these trends are not immediately clear, due to the large number of potentially significant processes occurring even in the absence of a phonon environment. This calls for future theoretical studies of effects such as phonon-induced transitions between laser-dressed states~\cite{Seidelmann2023_const_drive} and the sensitivity to Stark shifting of exciton states when photon pair emission is via the direct two-photon transition~\cite{Samal2024, Seidelmann2022_pulsed}.

\section{Conclusions}
\Red{We have used the numerically exact uniTEMPO method in combination with a degeneracy trick to model the non-Markovian dynamics of a system with large Hilbert space dimension.} We have applied this method to the problem of computing the entanglement of photon pairs emitted from the biexciton state of the quantum dot via cavity-enhanced two-photon emission, and subject to both non-Markovian phonon-induced decoherence and pulsed laser excitations. We have found that the approximate polaron master equation gives accurate results for the concurrence in the regimes that we have considered, thus significantly shortening the computational times of temporal behavior of % the calculation of trends within much more reasonable computer times. We find that 
photon entanglement, which decreases with both the duration of the laser pulse and the temperature of the phonon bath.

Future work will look at the optimal cavity coupling regime, in terms of the Purcell and quality factors of the cavity, which is a problem of significant experimental relevance. \Red{A further interesting extension of this analysis would be to explore the trade-offs between figures of merit such as concurrence, efficiency, indistinguishability and single-photon purity~\cite{Fischer2016, Iles-Smith2017, Gustin2018, Bauch2021, Bauch2023, Bracht2023, Vannucci2023, Heinisch2026}, the latter two of which are not easily computed using a polaron master equation method due to the necessary non-Markovian extension to the quantum regression theorem that is used to calculate two-time correlations~\cite{McCutcheon2016, Cosacchi2021}. We also emphasise the need to develop protocols that can exploit this unconventional source of hyperentangled photon pairs~\cite{Simon2005}.}

\begin{acknowledgments}
We thank Hamidreza Siampour for helpful discussions. K.S. acknowledges the UK EPSRC and SFI Centre for Doctoral Training in Photonic Integration and Advanced Data Storage (PIADS) program for the sponsorship of PhD studentship (grant no: EP/S023321/1). M.P. acknowledges support from the European Union’s Horizon Europe EIC-Pathfinder
project QuCoM (101046973), the Department for the Economy of Northern Ireland under the US-Ireland R\&D Partnership Programme, the ``Italian National Quantum Science and Technology Institute (NQSTI)" (PE0000023) - SPOKE 2 through project ASpEQCt.
\end{acknowledgments}

\appendix

\section{Spectral Density of the Phonon Environment}
\label{app:sd}
The dot-phonon coupling spectral density is given by
\begin{equation}
J(\omega) =
\frac{\omega^3}{4\pi^2 \rho \hbar c_s^5}
\left(
D_e e^{-\omega^2 a_e^2/(4c_s^2)}
-
D_h e^{-\omega^2 a_h^2/(4c_s^2)}
\right)^2,
\end{equation}
with  $D_e = 7.0~\mathrm{eV}$,
$D_h = -3.5~\mathrm{eV}$,
$\rho = 5370~\mathrm{kg/m^3}$,
$c_s = 5110~\mathrm{m/s}$,
$a_e/a_h = 1.15$, and
$a_e = 3.0~\mathrm{nm}$, which are  typical values for GaAs-based self-assembled quantum dots~\cite{Krummheuer2005}.

\section{\Red{Process Tensor Construction}}
\label{app:tempo}
\Red{The process tensor is computed using a discretized Feynman-Vernon path-integral, as follows
\begin{equation}
\mathcal{F}^{\mu_n,\ldots,\mu_1}
=
\prod_{j=1}^{n}
\prod_{l=0}^{\,n-j}
\left[b_l\right]^{\mu_{j+l},\mu_j},
\label{eq:3}
\end{equation}
where the elementary tensors $b_k$ depend on the eigenvalues $\lambda$ of the environmental coupling operator through}
\begin{equation}
    \Red{[b_k]^{\mu \nu} = 
    e^{-\left(\lambda_{\mu_l}-\lambda_{\mu_r}\right)
    \left(\eta_k \lambda_{\nu_l} - 
    \eta_k^* \lambda_{\nu_r}\right)}
    =
    \,}
    \begin{tikzpicture}[baseline=-2pt]
        \def\leg{0.2cm}      % vertical leg length
        \def\sep{0.25cm}   % horizontal separation between legs
    
        % Tensor node (rectangle)
        \node[
            draw=blue!60,
            fill=blue!10,
            line width=0.8pt,
            minimum width=0.75cm,
            minimum height=0.35cm
        ] (K) {$k$};
    
        % Bottom legs
        \draw ([xshift=-\sep]K.south) -- ++(0,-\leg) node[below] {$\mu$};
        \draw ([xshift=\sep]K.south)  -- ++(0,-\leg) node[below] {$\nu$};
    \end{tikzpicture}
\label{eq:b_k}
\end{equation}
%\vspace{0.2cm}
\Red{with the $\eta_k$ being determined by the bath correlation function $C(t)$,}
\begin{equation}
\eta_{i-j} =
\begin{cases}
\displaystyle
\int_{t_{i-1}}^{t_i} dt' 
\int_{t_{i-1}}^{t'} dt'' \,
C(t' - t'')
& i = j, \\[10pt]
\displaystyle
\int_{t_{i-1}}^{t_i} dt' 
\int_{t_{j-1}}^{t_j} dt'' \,
C(t' - t'')
& i \ne j,
\end{cases}
\end{equation}
where
\begin{equation}
C(t) =
\int_0^\infty d\omega \, J(\omega)
\left[
\coth\!\left(\frac{\beta \hbar \omega}{2} \right)
\cos(\omega t)
- i \sin(\omega t)
\right]
\end{equation}
and $\beta$ is the inverse temperature of the bath.

\Red{The product described by Eq.~\eqref{eq:3} can be depicted graphically as a tensor network, as shown in Fig.~\ref{fig:2}, by defining the tensors}
\begin{equation}
\Red{\delta_{ij}\delta^{\mu\nu} [b_k]^\mu_j
=
\,}
\begin{tikzpicture}[baseline=-2pt]
    \def\leg{0.2cm}      % vertical leg length
    \def\sep{0.25cm}   % horizontal separation between legs

    % Tensor node (rectangle)
    \node[
        draw=blue!60,
        fill=blue!10,
        line width=0.8pt,
        minimum width=0.75cm,
        minimum height=0.35cm
    ] (K) {$k$};

    % Top legs
    \draw ([xshift=-\sep]K.north) -- ++(0,\leg) node[above] {$j$};
    \draw ([xshift=\sep]K.north)  -- ++(0,\leg) node[above] {$\nu$};

    % Bottom legs
    \draw ([xshift=-\sep]K.south) -- ++(0,-\leg) node[below] {$\mu$};
    \draw ([xshift=\sep]K.south)  -- ++(0,-\leg) node[below] {$i$};
\end{tikzpicture}
\Red{\quad \mathrm{and} \quad
\delta^{\mu\nu} [b_0]^\mu_j
=
\,}
\begin{tikzpicture}[baseline=-2pt]
    \def\leg{0.2cm}      % vertical leg length
    \def\sep{0.25cm}   % horizontal separation between legs

    % Tensor node (rectangle)
    \node[
        draw=blue!60,
        fill=blue!10,
        line width=0.8pt,
        minimum width=0.75cm,
        minimum height=0.35cm
    ] (K) {$0$};

    % Top legs
    \draw ([xshift=-\sep]K.north) -- ++(0,\leg) node[above] {$j$};
    \draw ([xshift=\sep]K.north)  -- ++(0,\leg) node[above] {$\nu$};

    % Bottom legs
    \draw ([xshift=-\sep]K.south) -- ++(0,-\leg) node[below] {$\mu$};
\end{tikzpicture}
.
\end{equation}
\Red{We use the uniTEMPO algorithm~\cite{Link2024} for contracting and compressing this tensor network, which results in the uniform process tensor shown in Fig.~\ref{fig:2}.}

\Red{We now consider a coupling operator $\hat{O}$ with eigenvalues $\bar{\lambda}_{\bar{l}}$, corresponding to Liouville indices $\bar{\mu} = (\bar{\mu_l}, \bar{\mu_r})$, that have some degeneracy. We group together degenerate eigenvalues by relabelling $\bar{\mu}$ as $\bar{\mu} = (\mu, m) = ((\mu_l, \mu_r), (m_l, m_r))$ such that the eigenvalues satisfy $\bar{\lambda}_{\bar{\mu_l}} = \bar{\lambda}_{(\mu_l, m_l)} = \lambda_{\mu_l},$ i.e. they are degenerate in the second index. This relabelling allows us to rewrite the $[\bar{b}_k]^{\bar{\mu} \bar{\nu}}$ tensors as }
\begin{equation}
\begin{aligned}
    \Red{[\bar{b}_k]^{\bar{\mu} \bar{\nu}}} & \Red{= [\bar{b}_k]^{(\mu, m) (\nu, n)}} && \Red{= [b_k]^{\mu \nu} \, (\mathbf{1})^m (\mathbf{1})^n}\\
    & {}\\
    \begin{tikzpicture}[baseline=-2pt]
        \def\leg{0.2cm}      % vertical leg length
        \def\sep{0.25cm}   % horizontal separation between legs
    
        % Tensor node (rectangle)
        \node[
            draw=blue!60,
            fill=blue!10,
            line width=0.8pt,
            minimum width=0.75cm,
            minimum height=0.35cm
        ] (K) {$k$};
    
        % Bottom legs
        \draw[ultra thick] ([xshift=-\sep]K.south) -- ++(0,-\leg) node[below] {$\bar{\mu}$};
        \draw[ultra thick] ([xshift=\sep]K.south)  -- ++(0,-\leg) node[below] {$\bar{\nu}$};
    \end{tikzpicture}
    & = 
    \begin{tikzpicture}[baseline=-2pt]
        \def\leg{0.2cm}      % vertical leg length
        \def\sep{0.55cm}   % horizontal separation between
        \def\sepmum{0.25cm}
    
        % Tensor node (rectangle)
        \node[
            draw=blue!60,
            fill=blue!10,
            line width=0.8pt,
            minimum width=1.25cm,
            minimum height=0.35cm
        ] (K) {$k$};
    
        % Bottom legs
        \draw ([xshift=-\sep]K.south) -- ++(0,-\leg) node[below] {$\mu$};
        \draw[gray] ([xshift=-\sep+\sepmum]K.south) -- ++(0,-\leg) node[below] {$m$};
        \draw ([xshift=\sep-\sepmum]K.south)  -- ++(0,-\leg) node[below] {$\nu$};
        \draw[gray] ([xshift=\sep]K.south)  -- ++(0,-\leg) node[below] {$n$};
    \end{tikzpicture}
    && = 
    \begin{tikzpicture}[baseline=-2pt]
        \def\leg{0.2cm}      % vertical leg length
        \def\sep{0.25cm}   % horizontal separation between legs
    
        % Tensor node (rectangle)
        \node[
            draw=blue!60,
            fill=blue!10,
            line width=0.8pt,
            minimum width=0.75cm,
            minimum height=0.35cm
        ] (K) {$k$};
    
        % Bottom legs
        \draw ([xshift=-\sep]K.south) -- ++(0,-\leg) node[below] {$\mu$};
        \draw ([xshift=\sep]K.south)  -- ++(0,-\leg) node[below] {$\nu$};
    \end{tikzpicture}
    \begin{tikzpicture}[baseline=-2pt]
        \def\leg{0.35cm}
        \node[circle,
            draw=gray,
            fill=gray,
            line width=0.8pt,
            minimum size=0.15cm, inner sep=0pt
        ] (K){};
        % Bottom leg
        \draw[gray] (K.south) -- ++(0,-\leg) node[below] {$m$};
    \end{tikzpicture}
    \begin{tikzpicture}[baseline=-2pt]
        \def\leg{0.35cm}
        \node[circle,
            draw=gray,
            fill=gray,
            line width=0.8pt,
            minimum size=0.15cm, inner sep=0pt
        ] (K){};
        % Bottom leg
        \draw[gray] (K.south) -- ++(0,-\leg) node[below] {$n$};
    \end{tikzpicture}
    \end{aligned}
\end{equation}
\Red{where $(\mathbf{1})^i = 1$ for all $i$. Since the process tensor $\mathcal{F}$ is composed only of $b$ tensors, it factorises in just the same way, allowing one to calculate only the minimal process tensor for the set of non-degenerate eigenvalues, and use this to propagate the system state, as depicted graphically in the second row of Fig.~\ref{fig:2}.}

\section{Details of the Polaron Master Equation}
\label{app:pme}
The unitary polaron transform is defined as 
\begin{equation}
    H' = e^S H e^{-S}
\end{equation}
where $S=\hat{O}\,\sum_{k} \,g_k/\omega_k \, (b_k^\dagger-b_k)$.

In order to express the quantities entering the final master equation given in the main text, we first define the phonon propagator
\begin{equation}
\phi(t) =
\int_0^\infty d\omega \, \frac{J(\omega)}{\omega^2}
\left[
\coth\!\left(\frac{\beta \hbar \omega}{2} \right)
\cos(\omega t)
- i \sin(\omega t)
\right].
\end{equation}
The phonon-induced renormalization factor of the coupling constants $\langle B \rangle$ can then be expressed as $\langle B \rangle=e^{-\phi(0)/2}.$ The polaron shift is related to the spectral density through $\Delta_p = \int_0^\infty d\omega J(\omega)/\omega$, and the response functions $\kappa_{x/y}(\omega)$ are given by the Fourier transforms of the correlation functions $C_{x/y}(t)$ \cite{McCutcheon2011}
\begin{equation}
    \kappa_u(\omega) = \int_0^\infty dt \, C_u(t)e^{i\omega t}
\end{equation}
where \cite{Nazir2016}
\begin{equation}
\begin{aligned}
    C_x(t) &= \frac{\langle B \rangle^2}{2} 
    \left(e^{\phi(t)} + e^{-\phi(t)} - 2 \right)\\
    C_y(t) &= \frac{\langle B \rangle^2}{2} 
    \left(e^{\phi(t)} - e^{-\phi(t)} \right).
\end{aligned}
\end{equation}

\Red{
\section{Computational Resources}
\label{app:comp}
All calculations were performed on a conventional laptop computer equipped with Apple M2 Pro chip and 16 GB unified memory. The computation times associated with each data point in Fig.~\ref{fig:5} are typically 5 hours for uniTEMPO calculations and 50 minutes for PME calculations.}

\bibliographystyle{apsrev4-2}
\bibliography{references}

@article{McCutcheon2011,
  title = {A general approach to quantum dynamics using a variational master equation: Application to phonon-damped Rabi rotations in quantum dots},
  author = {McCutcheon, Dara P. S. and Dattani, Nikesh S. and Gauger, Erik M. and Lovett, Brendon W. and Nazir, Ahsan},
  journal = {Phys. Rev. B},
  volume = {84},
  issue = {8},
  pages = {081305},
  numpages = {4},
  year = {2011},
  month = {Aug},
  publisher = {American Physical Society},
  doi = {10.1103/PhysRevB.84.081305},
  url = {https://link.aps.org/doi/10.1103/PhysRevB.84.081305}
}

@article{Roy2011,
  title = {Influence of Electron--Acoustic-Phonon Scattering on Intensity Power Broadening in a Coherently Driven Quantum-Dot--Cavity System},
  author = {Roy, C. and Hughes, S.},
  journal = {Phys. Rev. X},
  volume = {1},
  issue = {2},
  pages = {021009},
  numpages = {19},
  year = {2011},
  month = {Nov},
  publisher = {American Physical Society},
  doi = {10.1103/PhysRevX.1.021009},
  url = {https://link.aps.org/doi/10.1103/PhysRevX.1.021009}
}

@article{Nazir2016,
doi = {10.1088/0953-8984/28/10/103002},
url = {https://doi.org/10.1088/0953-8984/28/10/103002},
year = {2016},
month = {feb},
publisher = {IOP Publishing},
volume = {28},
number = {10},
pages = {103002},
author = {Nazir, Ahsan and McCutcheon, Dara P S},
title = {Modelling exciton–phonon interactions in optically driven quantum dots},
journal = {Journal of Physics: Condensed Matter},
}

@article{Heinz2017,
  title = {Polarization-entangled twin photons from two-photon quantum-dot emission},
  author = {Heinze, Dirk and Zrenner, Artur and Schumacher, Stefan},
  journal = {Phys. Rev. B},
  volume = {95},
  issue = {24},
  pages = {245306},
  numpages = {9},
  year = {2017},
  month = {Jun},
  publisher = {American Physical Society},
  doi = {10.1103/PhysRevB.95.245306},
  url = {https://link.aps.org/doi/10.1103/PhysRevB.95.245306}
}

@misc{Dewan2026,
  author       = {Dewan, Urmimala and Kumar, Parvendra and Sarma, Amarendra K.},
  title        = {Polarization entanglement and qubit error rate dependence on the exciton-phonon coupling in self-assembled quantum dots},
  year         = {2026},
  eprint       = {2502.03413v4},
  archivePrefix= {arXiv},
  primaryClass = {quant-ph}
}

@article{Cosacchi2021,
  title = {Accuracy of the Quantum Regression Theorem for Photon Emission from a Quantum Dot},
  author = {Cosacchi, M. and Seidelmann, T. and Cygorek, M. and Vagov, A. and Reiter, D. E. and Axt, V. M.},
  journal = {Phys. Rev. Lett.},
  volume = {127},
  issue = {10},
  pages = {100402},
  numpages = {8},
  year = {2021},
  month = {Aug},
  publisher = {American Physical Society},
  doi = {10.1103/PhysRevLett.127.100402},
  url = {https://link.aps.org/doi/10.1103/PhysRevLett.127.100402}
}

@article{Cygorek2018_biex,
  title = {Comparison of different concurrences characterizing photon pairs generated in the biexciton cascade in quantum dots coupled to microcavities},
  author = {Cygorek, M. and Ungar, F. and Seidelmann, T. and Barth, A. M. and Vagov, A. and Axt, V. M. and Kuhn, T.},
  journal = {Phys. Rev. B},
  volume = {98},
  issue = {4},
  pages = {045303},
  numpages = {17},
  year = {2018},
  month = {Jul},
  publisher = {American Physical Society},
  doi = {10.1103/PhysRevB.98.045303},
  url = {https://link.aps.org/doi/10.1103/PhysRevB.98.045303}
}

@article{Seidelmann2023_pulsed_phonon,
  title = {Two-photon excitation with finite pulses unlocks pure dephasing-induced degradation of entangled photons emitted by quantum dots},
  author = {Seidelmann, T. and Bracht, T. K. and Lehner, B. U. and Schimpf, C. and Cosacchi, M. and Cygorek, M. and Vagov, A. and Rastelli, A. and Reiter, D. E. and Axt, V. M.},
  journal = {Phys. Rev. B},
  volume = {107},
  issue = {23},
  pages = {235304},
  numpages = {10},
  year = {2023},
  month = {Jun},
  publisher = {American Physical Society},
  doi = {10.1103/PhysRevB.107.235304},
  url = {https://link.aps.org/doi/10.1103/PhysRevB.107.235304}
}

@article{Glassl2012,
  title = {Impact of dark superpositions on the relaxation dynamics of an optically driven phonon-coupled exciton-biexciton quantum-dot system},
  author = {Gl\"assl, M. and Croitoru, M. D. and Vagov, A. and Axt, V. M. and Kuhn, T.},
  journal = {Phys. Rev. B},
  volume = {85},
  issue = {19},
  pages = {195306},
  numpages = {5},
  year = {2012},
  month = {May},
  publisher = {American Physical Society},
  doi = {10.1103/PhysRevB.85.195306},
  url = {https://link.aps.org/doi/10.1103/PhysRevB.85.195306}
}

@article{Liu2025, 
 title={Quantum correlations of spontaneous two-photon emission from a quantum dot}, volume={643}, DOI={https://doi.org/10.1038/s41586-025-09267-6}, number={8074}, journal={Nature}, publisher={Springer Science and Business Media LLC}, author={Liu, Shunfa and Wang, Yangpeng and Saleem, Yasser and Li, Xueshi and Liu, Hanqing and Yang, Cheng-Ao and Yang, Jiawei and Ni, Haiqiao and Niu, Zhichuan and Meng, Yun and Hu, Xiaolong and Yu, Ying and Wang, Xuehua and Cygorek, Moritz and Liu, Jin}, year={2025}, month={Jul}, pages={1234–1239}
}

@article{delValle2011,
doi = {10.1088/1367-2630/13/11/113014},
url = {https://doi.org/10.1088/1367-2630/13/11/113014},
year = {2011},
month = {nov},
publisher = {IOP Publishing},
volume = {13},
number = {11},
pages = {113014},
author = {del Valle, E and Gonzalez–Tudela, A and Cancellieri, E and Laussy, F P and Tejedor, C},
title = {Generation of a two-photon state from a quantum dot in a microcavity},
journal = {New Journal of Physics}
}

@article{Seidelmann2023_const_drive,
  title = {Phonon-induced transition between entangled and nonentangled photon emission in constantly driven quantum-dot--cavity systems},
  author = {Seidelmann, T. and Cosacchi, M. and Cygorek, M. and Reiter, D. E. and Vagov, A. and Axt, V. M.},
  journal = {Phys. Rev. B},
  volume = {107},
  issue = {7},
  pages = {075301},
  numpages = {13},
  year = {2023},
  month = {Feb},
  publisher = {American Physical Society},
  doi = {10.1103/PhysRevB.107.075301},
  url = {https://link.aps.org/doi/10.1103/PhysRevB.107.075301}
}

@article{Samal2024,
  title = {Effects of cavity-mediated processes on the polarization entanglement of photon pairs emitted from quantum dots},
  author = {Samal, Mukesh Kumar and Mishra, Divya and Kumar, Parvendra},
  journal = {Phys. Rev. A},
  volume = {109},
  issue = {1},
  pages = {013708},
  numpages = {8},
  year = {2024},
  month = {Jan},
  publisher = {American Physical Society},
  doi = {10.1103/PhysRevA.109.013708},
  url = {https://link.aps.org/doi/10.1103/PhysRevA.109.013708}
}

@article{Basset2023,
  title = {Signatures of the Optical Stark Effect on Entangled Photon Pairs from Resonantly Pumped Quantum Dots},
  author = {Basso Basset, F. and Rota, M. B. and Beccaceci, M. and Krieger, T. M. and Buchinger, Q. and Neuwirth, J. and Huet, H. and Stroj, S. and Covre da Silva, S. F. and Ronco, G. and Schimpf, C. and H\"ofling, S. and Huber-Loyola, T. and Rastelli, A. and Trotta, R.},
  journal = {Phys. Rev. Lett.},
  volume = {131},
  issue = {16},
  pages = {166901},
  numpages = {8},
  year = {2023},
  month = {Oct},
  publisher = {American Physical Society},
  doi = {10.1103/PhysRevLett.131.166901},
  url = {https://link.aps.org/doi/10.1103/PhysRevLett.131.166901}
}

@article{Seidelmann2022_pulsed,
  title = {Two-Photon Excitation Sets Limit to Entangled Photon Pair Generation from Quantum Emitters},
  author = {Seidelmann, T. and Schimpf, C. and Bracht, T. K. and Cosacchi, M. and Vagov, A. and Rastelli, A. and Reiter, D. E. and Axt, V. M.},
  journal = {Phys. Rev. Lett.},
  volume = {129},
  issue = {19},
  pages = {193604},
  numpages = {7},
  year = {2022},
  month = {Nov},
  publisher = {American Physical Society},
  doi = {10.1103/PhysRevLett.129.193604},
  url = {https://link.aps.org/doi/10.1103/PhysRevLett.129.193604}
}

@article{Pfanner2008,
  title = {Entangled photon sources based on semiconductor quantum dots: The role of pure dephasing},
  author = {Pfanner, Gernot and Seliger, Marek and Hohenester, Ulrich},
  journal = {Phys. Rev. B},
  volume = {78},
  issue = {19},
  pages = {195410},
  numpages = {8},
  year = {2008},
  month = {Nov},
  publisher = {American Physical Society},
  doi = {10.1103/PhysRevB.78.195410},
  url = {https://link.aps.org/doi/10.1103/PhysRevB.78.195410}
}

@article{delValle2010,
  title = {Two-photon lasing by a single quantum dot in a high-$Q$ microcavity},
  author = {del Valle, Elena and Zippilli, Stefano and Laussy, Fabrice P. and Gonzalez-Tudela, Alejandro and Morigi, Giovanna and Tejedor, Carlos},
  journal = {Phys. Rev. B},
  volume = {81},
  issue = {3},
  pages = {035302},
  numpages = {14},
  year = {2010},
  month = {Jan},
  publisher = {American Physical Society},
  doi = {10.1103/PhysRevB.81.035302},
  url = {https://link.aps.org/doi/10.1103/PhysRevB.81.035302}
}

@article{Pollock2018,
  title = {Non-Markovian quantum processes: Complete framework and efficient characterization},
  author = {Pollock, Felix A. and Rodr\'{\i}guez-Rosario, C\'esar and Frauenheim, Thomas and Paternostro, Mauro and Modi, Kavan},
  journal = {Phys. Rev. A},
  volume = {97},
  issue = {1},
  pages = {012127},
  numpages = {13},
  year = {2018},
  month = {Jan},
  publisher = {American Physical Society},
  doi = {10.1103/PhysRevA.97.012127},
  url = {https://link.aps.org/doi/10.1103/PhysRevA.97.012127}
}

@article{Strathearn2018, 
title={Efficient non-Markovian quantum dynamics using time-evolving matrix product operators}, volume={9}, 
url={https://doi.org/10.1038/s41467-018-05617-3}, 
number={1}, 
journal={Nature Communications}, author={Strathearn, A. and Kirton, P. and Kilda, D. and Keeling, J. and Lovett, B. W.}, year={2018}, 
pages={3322},
month={Aug} 
}

@article{Jorgensen2019,
  title = {Exploiting the Causal Tensor Network Structure of Quantum Processes to Efficiently Simulate Non-Markovian Path Integrals},
  author = {J\o{}rgensen, Mathias R. and Pollock, Felix A.},
  journal = {Phys. Rev. Lett.},
  volume = {123},
  issue = {24},
  pages = {240602},
  numpages = {7},
  year = {2019},
  month = {Dec},
  publisher = {American Physical Society},
  doi = {10.1103/PhysRevLett.123.240602},
  url = {https://link.aps.org/doi/10.1103/PhysRevLett.123.240602}
}

@article{Link2024,
  title = {Open Quantum System Dynamics from Infinite Tensor Network Contraction},
  author = {Link, Valentin and Tu, Hong-Hao and Strunz, Walter T.},
  journal = {Phys. Rev. Lett.},
  volume = {132},
  issue = {20},
  pages = {200403},
  numpages = {7},
  year = {2024},
  month = {May},
  publisher = {American Physical Society},
  doi = {10.1103/PhysRevLett.132.200403},
  url = {https://link.aps.org/doi/10.1103/PhysRevLett.132.200403}
}

@article{Cygorek2017,
  title = {Nonlinear cavity feeding and unconventional photon statistics in solid-state cavity QED revealed by many-level real-time path-integral calculations},
  author = {Cygorek, M. and Barth, A. M. and Ungar, F. and Vagov, A. and Axt, V. M.},
  journal = {Phys. Rev. B},
  volume = {96},
  issue = {20},
  pages = {201201},
  numpages = {5},
  year = {2017},
  month = {Nov},
  publisher = {American Physical Society},
  doi = {10.1103/PhysRevB.96.201201},
  url = {https://link.aps.org/doi/10.1103/PhysRevB.96.201201}
}

@article{Morreau2019,
  title = {Phonon-induced dephasing in quantum-dot--cavity QED},
  author = {Morreau, A. and Muljarov, E. A.},
  journal = {Phys. Rev. B},
  volume = {100},
  issue = {11},
  pages = {115309},
  numpages = {16},
  year = {2019},
  month = {Sep},
  publisher = {American Physical Society},
  doi = {10.1103/PhysRevB.100.115309},
  url = {https://link.aps.org/doi/10.1103/PhysRevB.100.115309}
}

@article{Hall2025,
  title = {Controlling dephasing of coupled qubits via shared bath coherence},
  author = {Hall, L. M. J. and Sirkina, L. S. and Morreau, A. and Langbein, W. and Muljarov, E. A.},
  journal = {Phys. Rev. B},
  volume = {112},
  issue = {4},
  pages = {045303},
  numpages = {16},
  year = {2025},
  month = {Jul},
  publisher = {American Physical Society},
  doi = {10.1103/ltk8-fpv3},
  url = {https://link.aps.org/doi/10.1103/ltk8-fpv3}
}

@article{Prior2010,
  title = {Efficient Simulation of Strong System-Environment Interactions},
  author = {Prior, Javier and Chin, Alex W. and Huelga, Susana F. and Plenio, Martin B.},
  journal = {Phys. Rev. Lett.},
  volume = {105},
  issue = {5},
  pages = {050404},
  numpages = {4},
  year = {2010},
  month = {Jul},
  publisher = {American Physical Society},
  doi = {10.1103/PhysRevLett.105.050404},
  url = {https://link.aps.org/doi/10.1103/PhysRevLett.105.050404}
}

@article{Chin2010,
	author = {Chin, Alex W. and Rivas, {\'A}ngel and Huelga, Susana F. and Plenio, Martin B.},
	journal = {Journal of Mathematical Physics},
	month = {09},
	number = {9},
	pages = {092109},
	title = {Exact mapping between system-reservoir quantum models and semi-infinite discrete chains using orthogonal polynomials},
	volume = {51},
	year = {2010}
}

@article{Makri1995a,
	author = {Makri, Nancy and Makarov, Dmitrii E.},
	journal = {The Journal of Chemical Physics},
	month = {03},
	number = {11},
	pages = {4600-4610},
	title = {Tensor propagator for iterative quantum time evolution of reduced density matrices. I. Theory},
	volume = {102},
	year = {1995}
}

@article{Makri1995b,
	author = {Makri, Nancy and Makarov, Dmitrii E.},
	journal = {The Journal of Chemical Physics},
	month = {03},
	number = {11},
	pages = {4611-4618},
	title = {Tensor propagator for iterative quantum time evolution of reduced density matrices. II. Numerical methodology},
	volume = {102},
	year = {1995}
}

@article{Laneve2025, title={Quantum teleportation with dissimilar quantum dots over a hybrid quantum network}, volume={16}, url={https://www.nature.com/articles/s41467-025-65911-9}, number={1}, journal={Nature Communications}, publisher={Springer Science and Business Media LLC}, author={Laneve, Alessandro and Ronco, Giuseppe and Beccaceci, Mattia and Barigelli, Paolo and Salusti, Francesco and Claro-Rodriguez, Nicolas and De Pascalis, Giorgio and Suprano, Alessia and Chiaudano, Leone and Schöll, Eva and Hanschke, Lukas and Krieger, Tobias M. and Buchinger, Quirin and Covre da Silva, Saimon F. and Neuwirth, Julia and Stroj, Sandra and Höfling, Sven and Huber-Loyola, Tobias and Usuga Castaneda, Mario A. and Carvacho, Gonzalo}, year={2025}, month={Nov}
}

@misc{Beccaceci2025arxiv,
  author       = {Mattia Beccaceci and Giuseppe Ronco and Fabrizio Cienzo and Pierpaolo Bassetti and Alessandro Laneve and Francesco Basso Basset and Tobias M. Krieger and Qurin Buchinger and Francesco Salusti and Barbara Souza Damasceno and Silke Kuhn and Saimon F. Covre da Silva and Sandra Stroj and Klaus D. Jöns and Sven Höfling and Tobias Huber-Loyola and Armando Rastelli and Michele B. Rota and Rinaldo Trotta},
  title        = {All-photonic entanglement swapping with remote quantum dots},
  year         = {2025},
  eprint       = {2512.10651},
  archivePrefix= {arXiv},
  primaryClass = {quant-ph},
}

@article{Iles-Smith2017, title={Phonon scattering inhibits simultaneous near-unity efficiency and indistinguishability in semiconductor single-photon sources}, volume={11}, DOI={https://doi.org/10.1038/nphoton.2017.101}, number={8}, journal={Nature Photonics}, author={Iles-Smith, Jake and McCutcheon, Dara P. S. and Nazir, Ahsan and Mørk, Jesper}, year={2017}, month={Jul}, pages={521–526}
}

@article{Azure2023,
  title = {Quantum repeaters: From quantum networks to the quantum internet},
  author = {Azuma, Koji and Economou, Sophia E. and Elkouss, David and Hilaire, Paul and Jiang, Liang and Lo, Hoi-Kwong and Tzitrin, Ilan},
  journal = {Rev. Mod. Phys.},
  volume = {95},
  issue = {4},
  pages = {045006},
  numpages = {66},
  year = {2023},
  month = {Dec},
  publisher = {American Physical Society},
  doi = {10.1103/RevModPhys.95.045006},
  url = {https://link.aps.org/doi/10.1103/RevModPhys.95.045006}
}

@article{Vannucci2023,
  title = {Highly efficient and indistinguishable single-photon sources via phonon-decoupled two-color excitation},
  author = {Vannucci, Luca and Gregersen, Niels},
  journal = {Phys. Rev. B},
  volume = {107},
  issue = {19},
  pages = {195306},
  numpages = {11},
  year = {2023},
  month = {May},
  publisher = {American Physical Society},
  doi = {10.1103/PhysRevB.107.195306},
  url = {https://link.aps.org/doi/10.1103/PhysRevB.107.195306}
}

@article{Krummheuer2005,
  title = {Pure dephasing and phonon dynamics in GaAs- and GaN-based quantum dot structures: Interplay between material parameters and geometry},
  author = {Krummheuer, B. and Axt, V. M. and Kuhn, T. and D'Amico, I. and Rossi, F.},
  journal = {Phys. Rev. B},
  volume = {71},
  issue = {23},
  pages = {235329},
  numpages = {13},
  year = {2005},
  month = {Jun},
  publisher = {American Physical Society},
  doi = {10.1103/PhysRevB.71.235329},
  url = {https://link.aps.org/doi/10.1103/PhysRevB.71.235329}
}

@article{McCutcheon2016,
  title = {Optical signatures of non-Markovian behavior in open quantum systems},
  author = {McCutcheon, Dara P. S.},
  journal = {Phys. Rev. A},
  volume = {93},
  issue = {2},
  pages = {022119},
  numpages = {7},
  year = {2016},
  month = {Feb},
  publisher = {American Physical Society},
  doi = {10.1103/PhysRevA.93.022119},
  url = {https://link.aps.org/doi/10.1103/PhysRevA.93.022119}
}

@article{Prilmuller2018,
  title = {Hyperentanglement of Photons Emitted by a Quantum Dot},
  author = {Prilm\"uller, Maximilian and Huber, Tobias and M\"uller, Markus and Michler, Peter and Weihs, Gregor and Predojevi\ifmmode \acute{c}\else \'{c}\fi{}, Ana},
  journal = {Phys. Rev. Lett.},
  volume = {121},
  issue = {11},
  pages = {110503},
  numpages = {5},
  year = {2018},
  month = {Sep},
  publisher = {American Physical Society},
  doi = {10.1103/PhysRevLett.121.110503},
  url = {https://link.aps.org/doi/10.1103/PhysRevLett.121.110503}
}

@article{Fischer2016, title={Dynamical modeling of pulsed two-photon interference}, volume={18}, DOI={https://doi.org/10.1088/1367-2630/18/11/113053}, number={11}, journal={New Journal of Physics}, author={Fischer, Kevin A and Müller, Kai and Lagoudakis, Konstantinos G and Vučković, Jelena}, year={2016}, month={Nov}, pages={113053}}

@article{Bracht2023,
author = {Thomas K. Bracht and Moritz Cygorek and Tim Seidelmann and Vollrath Martin Axt and Doris E. Reiter},
journal = {Optica Quantum},
number = {2},
pages = {103--107},
publisher = {Optica Publishing Group},
title = {Temperature-independent almost perfect photon entanglement from quantum dots via the SUPER scheme},
volume = {1},
month = {Dec},
year = {2023},
url = {https://opg.optica.org/opticaq/abstract.cfm?URI=opticaq-1-2-103},
doi = {10.1364/OPTICAQ.498559}
}

@article{Bauch2023, title={On‐Demand Indistinguishable and Entangled Photons Using Tailored Cavity Designs}, volume={7}, number={1}, journal={Advanced Quantum Technologies}, publisher={Wiley}, author={Bauch, David and Siebert, Dustin and Jöns, Klaus D. and Förstner, Jens and Schumacher, Stefan}, year={2023}, month={Nov} }

@article{Bauch2021,
  title = {Ultrafast electric control of cavity mediated single-photon and photon-pair generation with semiconductor quantum dots},
  author = {Bauch, David and Heinze, Dirk and F\"orstner, Jens and J\"ons, Klaus D. and Schumacher, Stefan},
  journal = {Phys. Rev. B},
  volume = {104},
  issue = {8},
  pages = {085308},
  numpages = {9},
  year = {2021},
  month = {Aug},
  publisher = {American Physical Society},
  doi = {10.1103/PhysRevB.104.085308},
  url = {https://link.aps.org/doi/10.1103/PhysRevB.104.085308}
}

@misc{Heinisch2026,
      title={High-quality single photons from cavity-enhanced biexciton-to-exciton transition}, 
      author={Nils Heinisch and Francesco Salusti and Mark R. Hogg and Timon L. Baltisberger and Malwina A. Marczak and Sascha R. Valentin and Arne Ludwig and Klaus D. Jöns and Richard J. Warburton and Stefan Schumacher},
      year={2026},
      eprint={2602.18153},
      archivePrefix={arXiv},
      primaryClass={quant-ph},
      url={https://arxiv.org/abs/2602.18153}, 
}

@article{Gustin2018,
  title = {Pulsed excitation dynamics in quantum-dot--cavity systems: Limits to optimizing the fidelity of on-demand single-photon sources},
  author = {Gustin, Chris and Hughes, Stephen},
  journal = {Phys. Rev. B},
  volume = {98},
  issue = {4},
  pages = {045309},
  numpages = {12},
  year = {2018},
  month = {Jul},
  publisher = {American Physical Society},
  doi = {10.1103/PhysRevB.98.045309},
  url = {https://link.aps.org/doi/10.1103/PhysRevB.98.045309}
}

@article{Scholl2020,
  title = {Crux of Using the Cascaded Emission of a Three-Level Quantum Ladder System to Generate Indistinguishable Photons},
  author = {Sch\"oll, Eva and Schweickert, Lucas and Hanschke, Lukas and Zeuner, Katharina D. and Sbresny, Friedrich and Lettner, Thomas and Trivedi, Rahul and Reindl, Marcus and Covre da Silva, Saimon Filipe and Trotta, Rinaldo and Finley, Jonathan J. and Vu\ifmmode \check{c}\else \v{c}\fi{}kovi\ifmmode \acute{c}\else \'{c}\fi{}, Jelena and M\"uller, Kai and Rastelli, Armando and Zwiller, Val and J\"ons, Klaus D.},
  journal = {Phys. Rev. Lett.},
  volume = {125},
  issue = {23},
  pages = {233605},
  numpages = {7},
  year = {2020},
  month = {Dec},
  publisher = {American Physical Society},
  doi = {10.1103/PhysRevLett.125.233605},
  url = {https://link.aps.org/doi/10.1103/PhysRevLett.125.233605}
}

@article{Strobel2025, title={Telecom-wavelength quantum teleportation using frequency-converted photons from remote quantum dots}, volume={16}, url={https://www.nature.com/articles/s41467-025-65912-8}, number={1}, journal={Nature Communications}, publisher={Springer Science and Business Media LLC}, author={Strobel, Tim and Vyvlecka, Michal and Neureuther, Ilenia and Bauer, Tobias and Schäfer, Marlon and Kazmaier, Stefan and Sharma, Nand Lal and Joos, Raphael and Weber, Jonas H. and Nawrath, Cornelius and Nie, Weijie and Bhayani, Ghata and Hopfmann, Caspar and Becher, Christoph and Michler, Peter and Portalupi, Simone Luca}, year={2025}, month={Nov} }

@article{Simon2005,
  title = {Creating Single Time-Bin-Entangled Photon Pairs},
  author = {Simon, Christoph and Poizat, Jean-Philippe},
  journal = {Phys. Rev. Lett.},
  volume = {94},
  issue = {3},
  pages = {030502},
  numpages = {4},
  year = {2005},
  month = {Jan},
  publisher = {American Physical Society},
  doi = {10.1103/PhysRevLett.94.030502},
  url = {https://link.aps.org/doi/10.1103/PhysRevLett.94.030502}
}

@article{delValle2013, title={Distilling one, two and entangled pairs of photons from a quantum dot with cavity QED effects and spectral filtering}, volume={15}, DOI={https://doi.org/10.1088/1367-2630/15/2/025019}, number={2}, journal={New Journal of Physics}, publisher={IOP Publishing}, author={del Valle, Elena}, year={2013}, month={Feb}, pages={025019–025019} }

@article{Cygorek2024-ACE,
	author = {Cygorek, Moritz and Gauger, Erik M.},
	doi = {10.1063/5.0221182},
	issn = {0021-9606},
	journal = {The Journal of Chemical Physics},
	month = {08},
	number = {7},
	pages = {074111},
	title = {ACE: A general-purpose non-Markovian open quantum systems simulation toolkit based on process tensors},
	url = {https://doi.org/10.1063/5.0221182},
	volume = {161},
	year = {2024}}

@article{Fux2024,
	author = {Fux, Gerald E. and Fowler-Wright, Piper and Beckles, Joel and Butler, Eoin P. and Eastham, Paul R. and Gribben, Dominic and Keeling, Jonathan and Kilda, Dainius and Kirton, Peter and Lawrence, Ewen D. C. and Lovett, Brendon W. and O'Neill, Eoin and Strathearn, Aidan and de Wit, Roosmarijn},
	journal = {The Journal of Chemical Physics},
	month = {09},
	number = {12},
	pages = {124108},
	title = {OQuPy: A Python package to efficiently simulate non-Markovian open quantum systems with process tensors},
	volume = {161},
	year = {2024}}

@article{Carmele2011,
  title = {Analytical solution of the quantum-state tomography of the biexciton cascade in semiconductor quantum dots: Pure dephasing does not affect entanglement},
  author = {Carmele, Alexander and Knorr, Andreas},
  journal = {Phys. Rev. B},
  volume = {84},
  issue = {7},
  pages = {075328},
  numpages = {5},
  year = {2011},
  month = {Aug},
  publisher = {American Physical Society},
  doi = {10.1103/PhysRevB.84.075328},
  url = {https://link.aps.org/doi/10.1103/PhysRevB.84.075328}
}

@article{McCutcheon2010,
doi = {10.1088/1367-2630/12/11/113042},
url = {https://doi.org/10.1088/1367-2630/12/11/113042},
year = {2010},
month = {nov},
publisher = {},
volume = {12},
number = {11},
pages = {113042},
author = {McCutcheon, Dara P S and Nazir, Ahsan},
title = {Quantum dot Rabi rotations beyond the weak exciton–phonon coupling regime},
journal = {New Journal of Physics}
}

\end{document}